\documentclass[12pt]{article}
\usepackage{graphicx}
\usepackage{amssymb,epsfig}
\usepackage{lscape,graphics,amsmath}
\usepackage{latexsym,amsfonts}
\usepackage{axodraw}
\usepackage[english]{babel}

\begin{document}

\begin{center}
{\Large \bf Quantum field-theoretical description of  neutrino and
neutral kaon oscillations
}\\ \vspace{4mm} Igor P.~Volobuev\\
\vspace{4mm} Skobeltsyn Institute of Nuclear Physics, Moscow State
University
\\ 119991 Moscow, Russia\\
\end{center}
\vspace{0.5cm}
\begin{flushright}
{ \textsl{Amicus Plato, sed magis amica est  veritas.\\
Aristotelis}}
\end{flushright}
\vspace{0.5cm}
\begin{abstract}
It is  shown that the neutrino and neutral kaon oscillation
processes can be consistently described in quantum field theory
using only the mass eigenstates of neutrinos and neutral kaons.
The  distance-dependent and time-dependent parts of the amplitudes
of these processes are calculated and the results turn out to be
in accord with those of the standard quantum mechanical
description of these processes based on the notion of neutrino
flavor states and neutral kaon states with definite strangeness.
However, the physical picture of the phenomena changes radically:
now there are no oscillations of flavor or definite strangeness
states, but,  instead of it, there is  an interference of
amplitudes due to different virtual mass eigenstates.
\end{abstract}

\section{Introduction}
Neutral kaon and neutrino oscillations are discussed in
theoretical physics for more than half a century already. The
standard theoretical approach to describing these phenomena is
based on considering the states with definite strangeness or
lepton number ($K^0, \bar K^0$ and the neutrino flavor states)
that are treated as superpositions of the corresponding mass
eigenstates. It is assumed that it is the former states that are
produced in the strong and  weak interactions. Considering their
time evolution in accordance with the Schr\"odinger equation, one
gets the oscillating probabilities of processes. This was first
demonstrated for neutral kaons in paper \cite{Pais:1955sm} and
later for neutrinos in papers
\cite{Pontecorvo:1957cp,Gribov:1968kq}. The present day status of
neutrino oscillations is discussed in detail in textbook
\cite{Bilenky:2010zza} and  in review article \cite{Petcov}.

However, although this approach to describing neutral kaon and
neutrino oscillations  is physically transparent and simple, it is
not rigorous and in many papers it was argued that it should be
considered as a phenomenological one (see, e.g.
\cite{Giunti:1993se,Grimus:1996av,Lobanov:2015esa,Naumov:2010um,Slad:2016axy}).
Its most noticeable flaw is the violation of the energy-momentum
conservation in the processes, in which $K^0, \bar K^0$ or a
neutrino flavor state are produced, because in local quantum field
theory, where the four-momentum is conserved in any interaction
vertex,  different mass-eigenstate components of these states must
have different momenta as well as different energies. Thus, if the
incoming particles   have definite four-momenta, then the produced
particles  must also have definite four-momenta, which implies
that neutral kaons and neutrinos can be produced only in the mass
eigenstates.  Apparently,  the spread of the momenta of the
incoming particles (the use of wave packets) does not solve the
problem, if the interaction remains local.

A simple physical idea for  solving the problem with the violation
of energy-momentum conservation is to go off the mass shell.
However, a specific realization of this idea encountered serious
difficulties.  Its first implementation in QFT was given in paper
\cite{Giunti:1993se}, where it was suggested that the produced
neutrino mass eigenstates are virtual  and their motion to the
detection point should be described by the Feynman propagators.
The authors managed to get through rather complicated calculations
and reproduced the results of the standard quantum mechanical
approach. Later this idea was developed in paper
\cite{Grimus:1996av}. The authors of this paper proved an
important theorem that, for large macroscopic distances between
the production and the detection points, the virtual neutrinos are
almost on the mass shell, and thus reduced the description in
terms of the virtual neutrinos to the standard one. Recently, the
approach with virtual neutrinos was also discussed in paper
\cite{Naumov:2010um}, where the authors used a formalism of
relativistic wave packets to facilitate the calculations.
Nevertheless, the calculations in all these papers are too bulky
and the final results are not quite convincing. This is due to the
standard S-matrix formalism of QFT used in these papers, which is
not appropriate for describing processes at finite distances and
finite time intervals.

In the present paper we will modify the standard perturbative
S-matrix formalism so as to be able to calculate the amplitudes of
the processes at finite distances and finite time intervals and
apply it to describing neutrino and neutral kaon oscillations. We
start with discussing neutrino oscillation processes, which is
simpler because in these processes  the massive  neutrinos can be
treated as stable particles \cite{Petcov:1976ff}.

\section{Neutrino oscillations}
We consider the  minimal extension of the Standard Model (SM) by
the right neutrino singlets. After the diagonalization of the
terms sesquilinear in the neutrino fields, the charged current
interaction Lagrangian of leptons takes the form

\begin{equation}\label{L_cc}
L_{cc} = - \frac{g }{2\sqrt{2}}\left(\sum_{i,k = 1}^3 \bar l_i
\gamma^\mu (1 - \gamma^5)U_{ik}\nu_k   W^{-}_\mu + h.c.\right),
\end{equation}
where $l_i$ denotes the field of the charged lepton of the i-th
generation, $\nu_i$ denotes the field of the neutrino mass
eigenstate most strongly coupled to $l_i$ and $U_{ik}$ stands for
the Pontecorvo-Maki-Nakagawa-Sakata (PMNS) matrix.

We consider the case, where the neutrinos are produced and
detected in the charged current interaction with  electrons.  Due
to the structure of the interaction Lagrangian any process
involving the production of a neutrino at one point and its
detection at another point, when treated perturbatively, includes
a subprocess described by the following diagram 
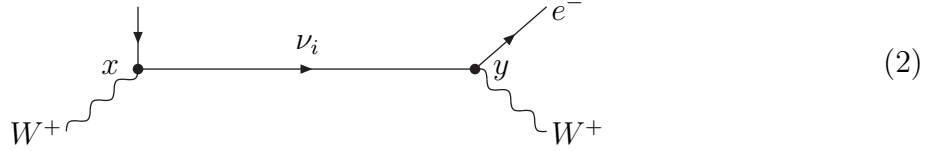
\vspace*{0.5cm}
\begin{center}
\begin{picture}(193,81)(0,0)
\ArrowLine(40.5,88.0)(40.5,64.5) \Text(33.5,65.5)[r]{$x$}
\Photon(13.5,41.0)(40.5,64.5){2}{3.0} \Text(12.0,41.0)[r]{$W^+$}
\Vertex (40.5,64.5){2} \ArrowLine(40.5,64.5)(167.5,64.5) \Vertex
(167.5,64.5){2} \Text(104.8,70.5)[b]{$\nu_i$}
\ArrowLine(167.5,64.5)(194.5,88.0) \Text(197.0,88.0)[l]{$e^-$}
\Text(175.0,64.5)[l]{$y$} \Photon(167.5,64.5)(194.5,41.0){2}{3.0}
\Text(197.0,41.0)[l]{$W^+$}
\Text(330.0,60.5)[b]{\addtocounter{equation}{1}(\arabic{equation})}
\label{diag}
\end{picture}
\end{center}
\vspace*{-1cm} for all the three neutrino mass eigenstates.
Depending on the neutrino production process, the external fermion
line at the point $x$ can either correspond to an incoming
electron or an outgoing positron. An important point is that there
is no need to add the diagram with interchanged W-bosons, because
they are associated with different spatially separated processes
and therefore should not be considered as identical particles.
Without loss of generality we can assume that the incoming
particles have definite momenta. Therefore all the three virtual
neutrino eigenstates and the outgoing particles also have definite
momenta.

The amplitude in the coordinate representation corresponding to
diagram (\ref{diag}) can be easily written out using the Feynman
rules formulated in \S 23 of textbook \cite{BOSH}. In the present
paper we will not calculate the complete amplitude, we will rather
calculate  only  its oscillating part that depends on the distance
between the points $x$ and $y$. This means that we can drop the
fermion wave functions that depend only on the particle momenta,
the polarization vectors of the W-bosons, as well as the coupling
constants and  the $\gamma$-matrix structures in the vertices.
What remains of the amplitude looks as follows:
\begin{equation}\label{amp_x}
|U_{1i}|^2 e^{-ipx} S^c_i(y - x) e^{iqy},
\end{equation}
where $S^c_i(y - x)$ is the Feynman propagator of the neutrino
mass eigenstate $\nu_i$, $p$ is the sum of the momenta of the
external lines at $x$ and $q$ is the sum of the momenta of the
external lines at $y$.

According to the prescriptions of the standard perturbative
S-matrix theory (\cite{BOSH}, \S 24), next we would have to
integrate  with respect to $x$ and $y$ over the Minkowski space,
which would give us the part of the scattering amplitude
corresponding to the inner fermion line of diagram (\ref{diag}).
However, in this case we would get the amplitude of the process
lasting an infinite amount of time and loose the information about
the distance between the production and  detection points. In
order to retain this information, we have to integrate with
respect to $x$ and $y$ in such a way that the distance between
these points along the direction of the neutrino propagation
remains fixed. Of course, this is at variance with the standard
S-matrix formalism. However, we recall that the diagram technique
in the coordinate representation was developed by R.~Feynman
\cite{Feynman:1949zx} without reference to S-matrix theory. Thus,
the Feynman diagrams in the coordinate representation make sense
beyond this theory, and for this reason we may integrate with
respect to $x$ and $y$ in any way depending on the physical
problem at hand. In particular, in the case under consideration we
have to integrate in such a way that the distance between the
points $x$ and $y$ along the direction of the neutrino propagation
equals to $L$. This can be achieved by introducing the delta
function $\delta(\vec p(\vec y -\vec x) - |\vec p|L)$ into the
integral, which gives the amplitude
\begin{equation}\label{amp_L}
|U_{1i}|^2 \int dx dy\, e^{-ipx} S^c_i(y - x)\, e^{iqy}\,
\delta(\vec p(\vec y -\vec x) - |\vec p|L).
\end{equation}

It is convenient to make the change of variables
\begin{equation}\label{chevar}
x = u - \frac{z}{2}, \quad y = u + \frac{z}{2}, \quad
\frac{D(x,y)}{D(u,z)} = 1
\end{equation}
and to integrate  with respect to $u$ in the integral in formula
(\ref{amp_L}). This transformation shows that the amplitude is
proportional to the delta function of energy-momentum conservation
$(2\pi)^4~\delta(q - p)$, which is unimportant for our further
considerations and will be dropped, and the distance-dependent
propagator of the neutrino mass eigenstate $\nu_i$ in the momentum
representation, which will be denoted by  $S^c_i(p,L)$ and is
defined by the remaining integral:
\begin{equation}\label{prop_L_mom}
S^c_i(p,L) = \int dz\, e^{ipz} S^c_i(z)\,  \delta(\vec p\vec z -
|\vec p|L).
\end{equation}
Obviously, the integration of this distance-dependent fermion
propagator with respect to $|\vec p|L$ from minus infinity to
infinity gives the standard Feynman fermion propagator in the
momentum representation. Thus, the distance-dependent fermion
propagator can be considered as the distribution of the Feynman
propagator with respect to $L$ along the direction of momentum
$\vec p$.

The integral in  formula (\ref{prop_L_mom}) is evaluated exactly
in Appendix 1, and for $\vec p^{\,2} > m_i^2 - p^2$ the result is
given by
\begin{equation}\label{prop_L_mom_a}
S^c_i(p,L) =  i\,\frac{\hat p + \vec \gamma \vec
 p\left(1 - \sqrt{1 + \frac{p^2 - m_i^2}{\vec p^{\,2}}}\,\right) + m_i }{2|\vec
p|\sqrt{\vec p^{\,2} + p^2 - m_i^2}}\, e^{-i\left(|\vec p| -
\sqrt{\vec p^{\,2} + p^2 - m_i^2}\,\right) L} \,.
\end{equation}
We emphasize that this distance-dependent fermion propagator makes
sense only for macroscopic distances. As we have already mentioned
in the Introduction,  the results of paper \cite{Grimus:1996av}
imply that the virtual particles propagating at macroscopic
distances are almost on the mass shell. This means that $|p^2 -
m_i^2|/ \vec p^{\,2} \ll 1$ and we can expand the square roots to
the first order in $(p^2 - m_i^2)/ \vec p^{\,2}$. It is clear that
this term can be dropped everywhere, except in the exponential,
where it is multiplied by a large macroscopic distance $L$, which
results in
\begin{equation}\label{prop_L_mom_bm}
S^c_i(p,L) = i\, \frac{\hat p + m_i }{ 2 \vec
p^{\,2}}\,e^{-i\frac{m_i^2 - p^2}{2|\vec p|} L}\,.
\end{equation}
It is worth noting that this distance-dependent fermion propagator
taken on the mass shell has no pole and does not depend on the
distance, which is also true  for the exact propagator in formula
(\ref{prop_L_mom_a}). The integration of this propagator with
respect to $|\vec p|L$ from zero to infinity gives one half of the
Feynman fermion propagator in the momentum representation (the
other half can be obtained by the integration of $S^c_i(p,-L)$
defined in accordance with formula (\ref{prop_L_mom})).

Since the neutrinos produced in the standard processes with
electrons or positrons are ultrarelativistic, the terms $m_i/|\vec
p|$ are negligibly small and we can also drop them. In this
approximation distance-dependent propagator (\ref{prop_L_mom_bm})
takes the simple form
\begin{equation}\label{prop_L_mom_b}
S^c_i(p,L) = i\, \frac{\hat p }{ 2 \vec p^{\,2}}\,e^{-i\frac{m_i^2
- p^2}{2|\vec p|} L}\,.
\end{equation}

Accordingly the amplitude of any process, where the neutrinos are
produced and detected in the charged current interaction with
electrons, is  the sum of the amplitudes with all the three
virtual neutrino mass eigenstates and contains,  due to
(\ref{amp_L}), (\ref{prop_L_mom}), (\ref{prop_L_mom_b}), the
scalar factor
\begin{equation}\label{amp_L_s}
\sum_{i = 1}^3|U_{1i}|^2\, e^{-i\frac{m_i^2 - p^2}{2|\vec p|}
L}\,.
\end{equation}
Therefore the probability of the process is proportional to
\begin{equation}\label{prob_L_s}
\left|\sum_{i = 1}^3|U_{1i}|^2\, e^{-i\frac{m_i^2 - p^2}{2|\vec
p|} L}\,\right|^2 = 1 - 4 \sum_{i,j = 1, i < j}^3|U_{1i}|^2
|U_{1j}|^2 \sin^2\left(\frac{m_j^2 - m_i^2}{4|\vec p|}\,L\right).
\end{equation}
In this way we have obtained the survival probability of the
electron, which coincides with the survival probability of the
electron neutrino, and  reproduced the result of the standard
quantum mechanical approach based on the notion of the neutrino
flavor states \cite{Giunti:2007ry}, although these states have not
been used in the derivation of formula (\ref{prob_L_s}).

Similarly, we can find the transition probability from the charged
lepton of the k-th generation to the charged lepton of the l-th
generation. The corresponding diagram looks like diagram
(\ref{diag}) with other incoming and outgoing fermions.  If the
momentum of the neutrino line satisfies the conditions $m_i/|\vec
p| \ll 1$, the amplitude of this process is proportional to
\begin{equation}\label{amp_L_s_kl}
\sum_{i = 1}^3\bar U_{ki}U_{li}\, e^{-i\frac{m_i^2 - p^2}{2|\vec
p|} L}\,,
\end{equation}
and one can easily obtain the general formulas for the oscillating
survival probabilities of charged leptons and transition
probabilities from any charged lepton to another charged lepton,
which again coincide with the corresponding standard formulas for
neutrino flavor states \cite{Giunti:2007ry}; we will not write
them out here.

Several remarks are in order. First, in the SM only the sum of all
three diagrams (\ref{diag}) with all the neutrino mass eigenstates
is gauge invariant, and therefore it is impossible to derive by
this method the formula for the oscillation of only two neutrino
flavors as it is usually done in the standard approach. Second, if
the neutrino production and detection processes are specified, one
can calculate the amplitude of the complete process by drawing the
corresponding Feynman diagram, constructing the amplitude in the
momentum space in accordance with the standard Feynman rules and
replacing the standard fermion propagator in the latter by the
distance-dependent propagator in formula (\ref{prop_L_mom_bm}). By
this method, one can also calculate the loop corrections to this
amplitude that will determine the actual accuracy of formula
(\ref{prob_L_s}). Third, to calculate the survival and transition
probabilities of charged leptons in realistic experiments one has
to generalize the approach by considering the amplitudes of
processes, where the distance between the production and detection
points lies between $L_1$ and $L_2$. Such an amplitude can be
obtained by integrating the amplitude in formula (\ref{amp_L})
with respect to $|\vec p|L$ from $|\vec p|L_1$ to $|\vec p|L_2$.
However, in the case $L_2 - L_1 \ll L_1$, which is specific for
experiments with neutrinos, the result is trivial: the amplitude
corresponding to the distance $L_1$ is just multiplied by $|\vec
p|(L_2 - L_1)$. We note once again that the integration of the
amplitude in formula (\ref{amp_L}) with respect to $|\vec p|L$
from minus infinity to infinity  gives the standard amplitude with
the  Feynman propagators. Thus,  the amplitude  integrated from
$|\vec p|L_1$ to $|\vec p|L_2$ can be considered as a portion of
the standard Feynman amplitude. Fourth, formula (\ref{prob_L_s})
is formally valid for any value of the macroscopic distance $L$.
The momentum spread of the produced neutrinos gives rise to the
standard coherence length, which limits its range of
applicability. In the framework of the present approach there may
exist one more coherence length arising due to a different reason.
As we have already noted, the virtual particles propagating at
macroscopic distances are almost on the mass shell. However, the
theorem proved in paper \cite{Grimus:1996av} does not define
explicitly the admissible deviation of the virtual particle
momentum squared from the mass shell depending on the distance
traveled. If such an admissible deviation exists, one can imagine
the situation, where, for certain large distances, the admissible
deviation from the mass shell is less than the minimal absolute
value of the differences of the neutrino masses squared. In this
case one should expect that either only two of the three diagrams
contribute to the amplitude of the process or only one of the
three diagrams contributes to the amplitude of the process, which
means that there would be either a change of the interference
pattern at a certain large distance or no interference at all
between the contributions of different neutrino mass eigenstates
and, therefore, no oscillation of the process probability
depending on the distance. For this reason it would be very
interesting to improve the theorem of paper \cite{Grimus:1996av}
so as to estimate the admissible deviation of virtual particles
from the mass shell depending on the distance.

\section{Neutral kaon oscillations}
Now we turn to discussing neutral kaon oscillations.  The
production of neutral kaons by 24~$GeV$ protons hitting a platinum
target and their subsequent decay to $\pi^+ \pi^-$ pairs were
studied in paper \cite{Geweniger:1973di}. It was shown in this
paper that the probability of detecting a $\pi^+ \pi^-$ pair
includes an oscillating component.

The standard theoretical explanation of this phenomenon is based
on the assumption that the state with strangeness -1 denoted $K^0$
is predominantly produced in the original interaction of protons
with platinum nuclei. This state has no definite mass and can be
treated as a superposition of states with definite masses. The
time evolution of this superposition of the mass eigenstates gives
rise to the oscillating decay probability to $\pi^+ \pi^-$ pairs
\cite{Pais:1955sm,Belusevic:1998pw}.

However, as we have already mentioned in the previous section, in
local QFT the production of a state without definite mass violates
the energy-momentum conservation. Below we will show, how this
problem can be solved by considering distance-dependent and
time-dependent amplitudes of  processes mediated by the particles
$K^0_S$ and $K^0_L$. The situation here is more complicated than
in the case of neutrinos for two reason: neutral kaons are states
of quark-antiquark pairs bound by the strong interaction and they
are unstable.

Let us denote the mass and the width of $K^0_S$ by $m_S, \Gamma_S$
and those of $K^0_L$ by $m_L, \Gamma_L$. Since these particles are
produced in the strong interaction processes, we cannot isolate
the subprocesses of $K^0_S, K^0_L$ production as definitely, as in
the case of only weakly interacting neutrinos. However, we can
assume that either a virtual $K^0_S$  or a virtual $K^0_L$ is
produced at a point $x$ and decays to $\pi^+ \pi^-$ pair at a
point $y$, which can be represented by the following diagram:
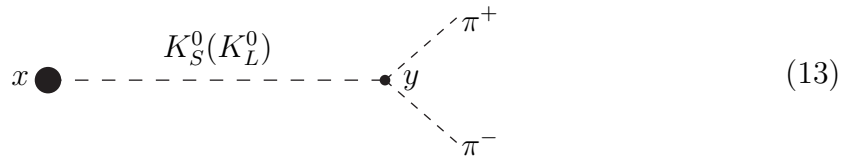
\vspace*{0.5cm}
\begin{center}
\begin{picture}(193,81)(0,0)
\Text(33.5,65.5)[r]{$x$}\Vertex (40.5,64.5){5}
\DashLine(40.5,64.5)(167.5,64.5){5} \Vertex (167.5,64.5){2}
\Text(104.8,70.5)[b]{$K^0_S (K^0_L)$}
\DashLine(167.5,64.5)(194.5,88.0){3} \Text(197.0,88.0)[l]{$\pi^+$}
\Text(175.0,64.5)[l]{$y$} \DashLine(167.5,64.5)(194.5,41.0){3}
\Text(197.0,41.0)[l]{$\pi^-$}
\Text(330.0,60.5)[b]{\addtocounter{equation}{1}(\arabic{equation})}
\label{diag_K}
\end{picture}
\end{center}
\vspace*{-1cm}  The amplitudes of the $\pi^+ \pi^-$ pair
production processes mediated by $K^0_S$ and $K^0_L$ differ in the
production amplitudes, the propagators and the decay amplitudes to
$\pi^+ \pi^-$. Denoting by  $P_S$ the production amplitude of
$K^0_S$  and by $A_{+-,S} $ its decay amplitude to $\pi^+ \pi^-$,
as well as  by  $P_L$ the production amplitude of $K^0_L$  and by
$A_{+-,L} $ its decay amplitude to $\pi^+ \pi^-$, we can write
down the amplitudes corresponding to diagram (10) as follows:
\begin{equation}\label{amp_x_K}
P_S\, e^{-ipx} D^c_S(x - y) e^{iqy}  A_{+-,S} ,\quad  P_L \,
e^{-ipx} D^c_L(x - y) e^{iqy} A_{+-,L},
\end{equation}
where $ D^c_S(x - y)$ and $ D^c_L(x - y)$ stand for the
propagators of $K^0_S$ and $K^0_L$, $p$ is the total momentum
entering the vertex at the point $x$ and $q$ is the sum of the
momenta of $\pi^+ \pi^-$ pair.

To find the distance-dependent amplitude of the process under
consideration, we again have to integrate these amplitudes  with
respect to $x$ and $y$  in such a way that the distance between
the points $x$ and $y$ along the direction of the neutral kaon
propagation equals to $L$. The amplitudes due to virtual $K^0_S$
and $K^0_L$ having absolutely the same form, we will do it in
detail only for the amplitude due to $K^0_S$ that looks like
\begin{equation}\label{amp_L_K}
P_S A_{+-,S}\,\int dx  dy\, e^{-ipx}D^c_S(x - y) e^{iqy} \,
\delta(\vec p(\vec y -\vec x) - |\vec p|L)\,.
\end{equation}
Again making the change of variables in accordance with
(\ref{chevar}), we obtain that the amplitude is proportional to
the delta function of energy-momentum conservation $(2\pi)^4
\delta(q -p)$, which is dropped, and the distance-dependent
propagator of  $K^0_S$ in the momentum representation that will be
denoted by $D^c_S(p,L)$ and is defined by the integral
\begin{equation}\label{prop_L_K}
D^c_S(p,L) = \int dz\, e^{ipz}D^c_S(z)  \, \delta(\vec p\vec z  -
|\vec p|L)\,.
\end{equation}
This integral is evaluated exactly in Appendix 2, the result for
$\vec p^{\,2} > m_S^2 - p^2 $ being given by
\begin{equation}\label{prop_L_mom_Ka}
D^c_S(p,L) =  \frac{i}{2|\vec p|\sqrt{\vec p^{\,2} + p^2 - m_S^2 +
i m_S \Gamma_S}}\,  e^{-i\left(|\vec p| - \sqrt{\vec p^{\,2} + p^2
- m_S^2 + i m_S \Gamma_S}\,\right) L} \,.
\end{equation}

We recall once more that the virtual particles propagating at
macroscopic distances are almost on the mass shell
\cite{Grimus:1996av}, and, therefore, $|p^2 - m_S^2|/ \vec p^{\,2}
\ll 1$. The ratio $m_S\Gamma_S/ \vec p^{\,2}$  being also very
small, we actually have  $|p^2 - m_S^2 + im_S\Gamma_S|/ \vec
p^{\,2} \ll 1$ and can drop the term $p^2 - m_S^2 + i m_S
\Gamma_S$ in the square root in the denominator and expand the
square root in the exponential to the first order in this term.
In this approximation formula (\ref{prop_L_mom_Ka}) for the
distance-dependent propagator of  $K^0_S$  can be written as
\begin{equation}\label{prop_L_mom_Kf}
D^c_S(p,L) =  \frac{i}{2\vec p^{\,2}}\,  e^{-i\frac{m_S^2 - p^2 -
i m_S \Gamma_S }{2|\vec p|} L} \,.
\end{equation}
The same calculations can be easily repeated for the
distance-dependent propagator of $K^0_L$, and from
(\ref{amp_x_K}), (\ref{amp_L_K}),(\ref{prop_L_mom_Kf}) one gets
that the total amplitude is proportional to
\begin{equation}\label{amp_L_mom_K_tot}
 e^{-\frac{ m_S \Gamma_S}{2|\vec p|} L} \, e^{-i\frac{ m_S^2 - p^2}
 {2|\vec p|} L}  + \frac{P_L }{P_S} \eta_{+-} e^{-\frac{ m_L
\Gamma_L}{2|\vec p|} L} \, e^{-i\frac{m_L^2 - p^2}{2|\vec p|}
L}\,,
\end{equation}
where $\eta_{+-} = A_{+-,L}/ A_{+-,S}$\, is the standard parameter
for describing the decay of neutral kaons to $\pi^+ \pi^-$ pairs
\cite{Belusevic:1998pw}. It is convenient to isolate the phases of
$\eta_{+-}$ and ${P_L }/{P_S}$ in accordance with the
parameterizations
\begin{equation}\label{phases}
\eta_{+-}  = |\eta_{+-}|e^{i\phi_{+-}}, \quad \frac{P_L }{P_S} =
\left|\frac{P_L }{P_S}\right| e^{i\phi_{P}}.
\end{equation}

 The specificity of the experiments with neutral
kaons suggests that their results should be represented not in
terms of the distance $L$, but rather in terms of  the time $T = L
p^0/|\vec p|$ \cite{Geweniger:1973di}. Substituting $L$ in terms
of $T$   into formula (\ref{amp_L_mom_K_tot}), we get
\begin{equation}\label{amp_LT_mom_K_tot}
e^{-\frac{ m_S \Gamma_S}{2p^0} \,T} \, e^{-i\frac{m_S^2 -
p^2}{2p^0} \,T} + \frac{P_L }{P_S} \eta_{+-} e^{-\frac{ m_L
\Gamma_L}{2p^0} \,T} \, e^{-i\frac{m_L^2 - p^2}{2p^0} \,T} \,.
\end{equation}

However, it is clear that  time-dependent amplitudes can be
obtained directly from amplitudes (\ref{amp_x_K}) by multiplying
them by the delta function $\delta(\beta(y^0 - x^0 - T))$ and then
integrating with respect to $x$ and $y$. Here  the parameter
$\beta$ has dimension of mass and will be defined later. Similar
to the case of the distance-dependent amplitude, the integral
entering the expression for the time-dependent amplitude due to
virtual $K^0_S$ can be represented as the product of the delta
function of energy-momentum conservation and  the time-dependent
propagator of $K^0_S$ in the momentum representation $D^c_S(p,T)$
defined by the formula
\begin{equation}\label{prop_T_K}
D^c_S(p,T) = \int dz \, e^{ipz}D^c_S(z)  \, \delta(\beta (z^0 -
T))\,.
\end{equation}
The integral in this formula is evaluated exactly in Appendix 3,
and the resulting time-dependent propagator looks like
\begin{equation}\label{prop_T_mom_4_K_S}
D^c_S(p,T) = \frac{i}{2\beta\sqrt{(p^0)^2 + m_S^2 -  p^2 - i m_S
\Gamma_S}}\, e^{i\left(p^0 -\sqrt{(p^0)^2 + m_S^2 - p^2 - i m_S
\Gamma_S}\,\right) T} \,.
\end{equation}
Again we have  $| m_S^2 - p^2 - im_S\Gamma_S|/  (p^0)^2 \ll 1$ and
for this reason we can drop the term $m_S^2 - p^2 - im_S\Gamma_S$
in the square root in the denominator and expand the square root
in the exponential to the first order in this term.  In this
approximation formula (\ref{prop_T_mom_4_K_S}) takes the form
\begin{equation}\label{prop_T_mom_Kf}
D^c_S(p,T) =  \frac{i}{2\beta p^{0}}\, e^{-i\frac{m_S^2 - p^2 -
im_S \Gamma_S}{2p^0} \,T}\,,
\end{equation}
which coincides with (\ref{prop_L_mom_Kf}), if we choose $\beta =
\vec p^{\,2}/p^0$ and substitute $T = L p^0/ |\vec p|$. Thus,  if
one uses the matched delta functions in the propagator
definitions, the distance-dependent propagator and the
time-dependent propagator at macroscopic distances and time
intervals are, in fact, the same and differ only in the
parametrization. It is worth noting that this is also true for the
case of neutrinos, where formula (\ref{prop_L_mom_b}) can be
obtained by substituting $\beta = \vec p^{\,2}/p^0, \, T = L p^0/
|\vec p|$\,\, into the corresponding time-dependent propagator.
Moreover, in the case of ultrarelativistic neutrinos one can put
$\beta = p^0$ and get the time-dependent propagator with the same
accuracy.

Summing the time-dependent amplitudes due to  $K^0_S$ and $K^0_L$,
one again gets that the total amplitude is proportional to the
expression in formula (\ref{amp_LT_mom_K_tot}). It is customary to
write this formula in terms of the proper time, which, in the case
under consideration, should be defined as follows:
\begin{equation}\label{p_time}
t_p = \frac{2 p^0 \,T}{m_S + m_L}.
\end{equation}
Formula (\ref{amp_LT_mom_K_tot}) rewritten in terms of the proper
time looks like
\begin{equation}\label{amp_t_p_mom_K_tot}
e^{-\frac{ m_S \Gamma_S}{m_S +\, m_L} \,t_p} \, e^{-i\frac{m_S^2 -
p^2 }{m_S +\, m_L} \,t_p} + {\left|\frac{P_L }{P_S}\right|
|\eta_{+-}|}\, e^{i(\phi_P + \phi_{+-})} e^{-\frac{ m_L
\Gamma_L}{m_S +\, m_L} \,t_p} \, e^{-i\frac{ m_L^2 - p^2}{m_S +\,
m_L} \,t_p} \,,
\end{equation}
where the phases $\phi_P$ and  $\phi_{+-}$  have been defined in
(\ref{phases}). Therefore the probability of the $\pi^+ \pi^-$
production process is proportional to
\begin{eqnarray}\nonumber
&&e^{-\frac{2 m_S \Gamma_S}{m_S +\, m_L} \,t_p} + {\left|\frac{P_L
}{P_S}\right|^2 |\eta_{+-}|^2 }\,  e^{-\frac{2 m_L \Gamma_L}{m_S
+\, m_L} \,t_p} + \\ \label{prob_t_K} &+&  2 \left|\frac{P_L
}{P_S}\right| |\eta_{+-}| e^{-\frac{ m_S \Gamma_S + \, m_L
\Gamma_L}{m_S +\, m_L} \,t_p} \cos\left( \Delta m_{L,S}\, t_p -
\phi_P - \phi_{+-} \right),
\end{eqnarray}
where $\Delta m_{L,S} = m_L - m_S $. For $P_L = P_S$ and $P_L =
-P_S$ this expression coincides exactly with those presented in
\cite{Belusevic:1998pw}, \S 4.3, if one neglects the difference
between $m_L$ and  $m_S$ in all the exponentials. This formula can
also be rewritten in the form
\begin{eqnarray}\nonumber
&& \left(e^{-\frac{ m_S \Gamma_S}{m_S +\, m_L} \,t_p} +
{\left|\frac{P_L }{P_S}\right| |\eta_{+-}| }\,  e^{-\frac{ m_L
\Gamma_L}{m_S +\, m_L} \,t_p}\right)^2 -  \\ \label{prob_t_K1} &-&
4 \left|\frac{P_L }{P_S}\right| |\eta_{+-}| e^{-\frac{ m_S
\Gamma_S + \, m_L \Gamma_L}{m_S +\, m_L} \,t_p} \sin^2\left(
\frac{\Delta m_{L,S}}{2}\, t_p - \frac{\phi_P + \phi_{+-}}{2}
\right),
\end{eqnarray}
which is most similar to formula (\ref{prob_L_s}) for neutrino
oscillations. Thus, we have again reproduced the results of the
standard approach without reference to the states with definite
strangeness $K^0, \bar K^0.$

\section{Conclusion}
In the present paper we have shown that it is possible to give a
consistent quantum field-theoretical description of  neutrino
oscillations in the SM minimally extended by the right neutrino
singlets. To this end we had just to adapt the standard
perturbative S-matrix formalism for calculating the amplitudes of
the processes passing at finite distances and finite time
intervals. The developed approach is physically transparent  and,
unlike  the standard one, has the advantage of not violating the
energy-momentum conservation. In its framework, the calculations
of amplitudes are much simpler than in papers
\cite{Giunti:1993se,Naumov:2010um}, where a similar approach based
on the standard S-matrix description of the propagation of virtual
neutrinos is used.

The application of this modified formalism to describing the
neutrino and neutral kaon oscillation processes showed that the
standard results can be easily and consistently obtained using
only the mass eigenstates of these particles.  Therefore, the
neutrino flavor states and the neutral kaon states with definite
strangeness are redundant in the theory and should be amputated by
Occam's razor.

Although the results obtained within the developed approach
coincide with the standard results of the neutrino and neutral
kaon oscillation theory, the physical picture of the  phenomena
changes radically. If there are no neutrino flavor states, there
is no neutrino oscillation, and the term can be retained only as a
historical one. What remains of this theory is the observable
oscillating survival probabilities of charged leptons  and the
observable oscillating transition probabilities from one charged
lepton to another charged lepton arising due to the interference
of the amplitudes of the processes mediated by different neutrino
mass eigenstates. Thus, it is the mass eigenstates $\nu_i, \, i =
1,2,3,$ that are the only observable neutrino states in the SM.
These states  can be consistently treated as particles, and it is
natural to call them after the charged leptons, to which they are
most strongly coupled, the electron neutrino, the muon neutrino
and the tau neutrino respectively, which restores the quark-lepton
symmetry in the SM.

In the present paper we have considered only the case of Dirac
neutrinos. However, all the results can be transferred, \emph{
mutatis mutandis}, to the case of Majorana neutrinos, because
their propagators have essentially the same structure.

A similar situation takes place in the case of neutral kaons,
where only the states $K^0_S$ and $K^0_L$ can be considered to be
particles. The states $K^0$ and $\bar K^0$ are artificial abstract
entities that are unnecessary in QFT and cannot exist or even be
produced in Nature, where, unlike in the theory, the weak
interaction cannot be switched off. Therefore, we have to admit
that there is no neutral kaon oscillations, but there is an
interference of the amplitudes with virtual $K^0_S$ and $K^0_L$
instead.

Finally, we note that one can construct the amplitudes and find
the cross sections of processes with any particles, whose
production and detection sites are separated by a macroscopic
distance, by drawing the corresponding Feynman diagrams,
constructing the amplitudes in the momentum space in accordance
with the standard Feynman rules and replacing the standard
propagators in the latter by the corresponding distance-dependent
or time-dependent propagators introduced in the present paper.
Such calculations may be useful for analyzing events in the
experiments, where detectors are situated at macroscopic distances
from the interaction points.

\bigskip
{\large \bf Acknowledgments}
\medskip \\
\noindent The author is grateful to E. Boos,  A. Lobanov and  M.
Smolyakov for reading the manuscript and making important
comments. The work was supported by grant NSh-7989.2016.2 of the
President of Russian Federation.

\setcounter{equation}{0}
\renewcommand{\theequation}{A\arabic{equation}}

\section*{Appendix 1}
Substituting  the standard integral representations for the
fermion propagator and the delta function  into the integral
defining the distance-dependent propagator in formula
(\ref{prop_L_mom}), we get
\begin{equation}\label{prop_L_mom_1}
S^c_i(p,L) = \frac{1}{(2\pi)^5} \int dz dk  dt\, e^{ipz} e^{-ikz}
e^{i(\vec p\vec z - |\vec p|L)t} \frac{\hat k + m_i }{m^2_i - k^2
- i\epsilon}\,,
\end{equation}
where $m_i$ denotes the mass of the neutrino mass eigenstate
$\nu_i$.

Next we make the change of variable $t = \omega/|\vec p|$ and
integrate with respect to $z$, which results in
\begin{equation}\label{prop_L_mom_2}
S^c_i(p,L) =  \frac{ 1}{2\pi |\vec p|} \int d\omega  dk \,
\delta(p^0 - k^0) \delta(\vec k - \vec p + \vec n \omega )
e^{-i\omega L} \frac{\hat k + m_i }{m^2_i - k^2 - i\epsilon}\,,
\end{equation}
where $\vec n = \vec p / |\vec p|$.

 The integration with respect
to $k$ is trivial due to the $\delta$-functions:
\begin{equation}\label{prop_L_mom_3}
S^c_i(p,L) =   \frac{ 1}{2\pi |\vec p|} \int d\omega \,
e^{-i\omega L} \frac{\hat p + \vec \gamma \vec n \omega + m_i
}{m^2_i - p^2  - 2  |\vec p| \omega + \omega^2 - i\epsilon}\,.
\end{equation}

The last integral can be evaluated for $ \vec p^{\,2} > m_i^2 -
p^2$ by closing the integration contour in the lower complex
half-plane and calculating the residue at the simple pole, which
gives:
\begin{equation}\label{prop_L_mom_4}
S^c_i(p,L) =  i\, \frac{\hat p + \vec \gamma \vec
 p\left(1 - \sqrt{1 + \frac{p^2 - m_i^2}{\vec p^{\,2}}}\right) + m_i }{2|\vec
p|\sqrt{\vec p^{\,2} + p^2 - m_i^2}}\, e^{-i\left(|\vec p| -
\sqrt{\vec p^{\,2} + p^2 - m_i^2}\,\right) L} \,.
\end{equation}

\section*{Appendix 2}
Again substituting  the standard integral representations for the
scalar field propagator and the delta function  into the integral
in formula (\ref{prop_L_K}), we get
\begin{equation}\label{prop_L_mom_K}
D^c_S(p,L) = \frac{1}{(2\pi)^5}\,  \int dz dk dt\, e^{ipz} e^{ikz}
e^{i(\vec p\vec z - |\vec p|L)t} \frac{1 }{m^2_S - k^2 - im_S
\Gamma_S}\,.
\end{equation}
Then we make the change of variable $t = \omega/|\vec p|$ and
integrate with respect to $z$ and $k$ similar to
(\ref{prop_L_mom_2}), (\ref{prop_L_mom_3}), which gives
\begin{equation}\label{prop_L_mom_3_K}
D^c_S(p,L) = \frac{1}{2\pi  |\vec p|} \int d\omega \, e^{-i\omega
L} \frac{1}{m^2_S - p^2  - 2  |\vec p| \omega + \omega^2 - im_S
\Gamma_S}\,.
\end{equation}

For $ \vec p^{\,2} > m_S^2 - p^2$, the  residue integration method
gives:
\begin{equation}\label{prop_L_mom_4_K}
D^c_S(p,L) = \frac{i}{2|\vec p|\sqrt{\vec p^{\,2} + p^2 - m_S^2 +
i m_S \Gamma_S}}\, e^{-i\left(|\vec p| -\sqrt{\vec p^{\,2} + p^2 -
m_S^2 + i m_S \Gamma_S}\,\right) L} \,.
\end{equation}

\section*{Appendix 3}
Once more substituting  the standard integral representations for
the scalar field propagator and the delta function  into the
integral in formula (\ref{prop_T_K}), we get
\begin{equation}\label{prop_T_mom_K}
D^c_S(p,T) = \frac{1}{(2\pi)^5}\,  \int dz  dk dt\, e^{ipz}
e^{ikz} e^{i\beta ( z^0  - T)t}\, \frac{1 }{m^2_S - k^2 - im_S
\Gamma_S}\,.
\end{equation}

To evaluate this integral  we make the change of variable $t =
\omega/\beta $, then  integrate with respect to  $z$ and $k$
similar to (\ref{prop_L_mom_2}), (\ref{prop_L_mom_3}) and arrive
at the result
\begin{equation}\label{prop_T_mom_3_K}
D^c_S(p,T) = \frac{1}{2\pi \beta} \int d\omega \, e^{-i\omega T}
\frac{1}{m^2_S - p^2  - 2   p^0 \omega - \omega^2 - im_S
\Gamma_S}\,.
\end{equation}

The  residue integration method gives:
\begin{equation}\label{prop_T_mom_4_K}
D^c_S(p,T) = \frac{i}{2\beta\sqrt{(p^0)^2 + m_S^2 -  p^2 - i m_S
\Gamma_S}}\, e^{i\left(p^0 -\sqrt{(p^0)^2 + m_S^2 -  p^2 - i m_S
\Gamma_S}\,\right) T} \,.
\end{equation}

\end{document}